\documentclass[sigconf,authorversion,nonacm]{acmart}

\AtBeginDocument{%
  }

\usepackage[noend]{algpseudocode}
\usepackage[ruled,vlined,linesnumbered]{algorithm2e}
\usepackage{algorithmicx}

\usepackage{multicol}

\usepackage{amsmath}

\SetKwInOut{Input}{Input}
\SetKwInOut{Output}{Output}

\begin{document}

\title{SAND: A Self-supervised and Adaptive NAS-Driven Framework for Hardware Trojan Detection}

\author{Zhixin Pan, Ziyu Shu, Linh Nguyen, and Amberbir Alemayoh}




\renewcommand{\shortauthors}{Trovato et al.}

\begin{abstract}
The globalized semiconductor supply chain has made Hardware Trojans (HT) a significant security threat to embedded systems, necessitating the design of efficient and adaptable detection mechanisms. 
Despite promising machine learning-based HT detection techniques in the literature, they suffer from ad hoc feature selection and the lack of adaptivity, all of which hinder their effectiveness across diverse HT attacks.
In this paper, we propose SAND, a self-supervised and adaptive NAS-driven framework for efficient HT detection. Specifically, this paper makes three key contributions. (1) We leverage self-supervised learning (SSL) to enable automated feature extraction, eliminating the dependency on manually engineered features. (2) SAND integrates neural architecture search (NAS) to dynamically optimize the downstream classifier, allowing for seamless adaptation to unseen benchmarks with minimal fine-tuning. (3) Experimental results show that SAND achieves a significant improvement in detection accuracy (up to 18.3\%) over state-of-the-art methods, exhibits high resilience against evasive Trojans, and demonstrates strong generalization.
\end{abstract}

\maketitle
\let\thefootnote\relax\footnotetext{This paper was published in the 2025 IEEE International Conference on Computer Design}

\section{Introduction}
\label{sec:Intro}

The rapid advancement of System-on-Chip (SoC) technology has led to increasing design scales, necessitating the globalization of the semiconductor supply chain to meet time-to-market constraints. However, the inevitable integration of third-party Intellectual Property (IP) vendors in the supply chain introduces security risks, exposing SoCs to Hardware Trojans (HTs) attacks. HTs pose severe threats, including information leakage and erroneous execution. Therefore, mitigating HT attacks is crucial to ensuring the trustworthiness of modern SoCs.

\textit{This paper focuses on trigger-activated HTs}, which are the predominant type in current literature. Their trigger conditions are often designed using a combination of rare events (signals, transitions) to evade detection, posing the efficient defense against such HT attacks a serious challenge.

Conventional HT detection methods fall into two main categories: {\it formal verification} and {\it test generation}. Formal verification methods aim to mathematically prove the correctness of a circuit by exhaustively checking predefined security properties through a SAT solver~\cite{nahiyan2017hardware}. However, they are often impractical for large-scale SoCs due to the high computational cost caused by exponential time complexity. 
Test generation methods, instead, apply carefully crafted test vectors to trigger anomalous behaviors or side-channel information, indicating the presence of HTs~\cite{huang2021machine}. While effective performance was demonstrated in existing works, they are time-consuming to detect well-hidden Trojans~\cite{witharana2021directed}.

Given the above limitations, machine learning (ML) has emerged as a promising approach for HT detection, demonstrating superior performance over traditional techniques. However, existing ML-based approaches also suffer from several critical limitations. First, these methods depend largely on manually selected feature vectors without a standardized guideline. Additionally, most ML models exhibit poor generalizability when applied to unseen variants of HTs, necessitating expensive efforts for model retraining~\cite{huang2020survey}. This is especially problematic in the context of HTs detection, as Trojans may vary significantly across different devices, thus requiring frequent model updates. 

To address these challenges, we propose SAND, a Self-supervised and Adaptive Detector for HT. The proposed framework enables an efficient integration between self-supervised learning (SSL) and neural architecture search (NAS). Specifically, the proposed work offers three major contributions:

\begin{itemize}
    \item {\bf Automated Feature Embedding via SSL}: Unlike existing works relying on manually selected features, SAND employs SSL to pre-train an encoder that captures meaningful representations from hardware benchmarks.
    \item {\bf Adaptive Classifier Optimization via NAS}: SAND leverages NAS to dynamically search for an optimal neural network architecture tailored to specific detection tasks with minimum retraining overhead.
    \item {\bf Comprehensive Experimental Evaluation}: Experimental results show that SAND achieves a significant improvement in detection accuracy (up to 18.3\%), adaptivity, and stability over state-of-the-art methods.
\end{itemize}



\section{Related Work and Background}
\label{sec:bgd}


\subsection{ML-based HT Detection}
\label{sec:MLHT}
As discussed in Section~\ref{sec:Intro}, ML-based HT detection approaches have demonstrated superior performance over traditional techniques. The majority of these methods rely on supervised learning~\cite{nasteski2017overview}, where the ML model is trained to identify HTs through pattern recognition of pre-defined circuit features. For example, a support vector machine (SVM)-based model was proposed in~\cite{gubbi2023hardware} to detect anomalous behaviors associated with HT activations in simulation-based environments. Similarly, deep learning-based techniques have been employed for HT detection in~\cite{yu2021deep}. Other ML models, such as random forests~\cite{kurihara2021hardware}, convolutional neural networks (CNNs)\cite{yasaei2022golden}, and reinforcement learning\cite{pan2021automated,chen2023adatest}, have also been explored, obtaining promising classification accuracy. Despite these advancements, ML-based HT detection still suffers from critical limitations. First, most existing methods rely on {\bf manually selected features} based on human expert knowledge, without a universally applicable strategy. For example, Hasegawa et al.~\cite{hasegawa2017hardware} identified five key features from gate-level netlists to train a classifier for HT detection. Similarly, Zhou et al.~\cite{zhou2016novel} developed a feature extraction algorithm by analyzing the distribution of rare signals within IP cores. While effective within specific contexts, these domain-specific methods struggle to be applied universally.



Second, these methods often exhibit {\bf poor generalizability}, struggling to detect unseen HT variants and requiring frequent retraining~\cite{huang2020survey}. 
According to a recent study~\cite{gaber2024malware}, the detection performance of several existing algorithms fluctuates across different benchmarks, often dropping below 60\%, which is no better than a random guess.
This highlights the urgent need for more efficient and adaptive HT detection frameworks.


\subsection{Self-Supervised Learning (SSL)}
\label{SSL}

The limitations of existing ML-based HT detection methods, particularly their reliance on manually designed feature representations, have highlighted the need for more adaptive approaches. A promising direction to address this challenge is through {\bf self-supervised learning (SSL)}, which has demonstrated remarkable success in automated feature embedding.

\begin{figure}[htp]
    \centering
    \includegraphics[width=0.9\linewidth]{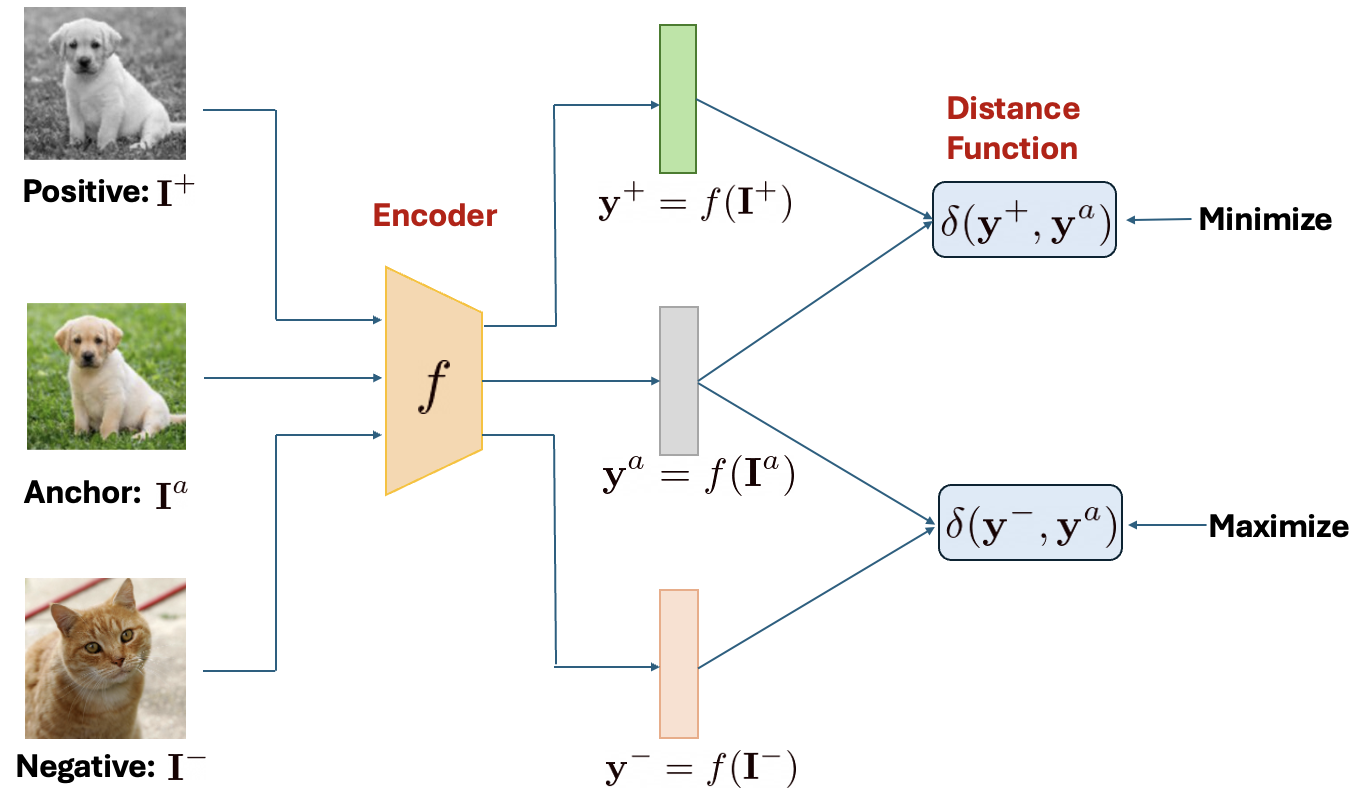}
    \vspace{-0.1in}
    \caption{Illustration of contrastive learning. Given an anchor input \( I^a \), a positive example \( I^+ \) is generated through data augmentation, while a negative example \( I^- \) is selected from a different class. The model learns a feature representation such that the distance \( \delta(I^a, I^+) \) is minimized, while the distance \( \delta(I^a, I^-) \) is maximized.}
    \vspace{-0.1in}
    \label{fig:contrastive_learning}
\end{figure}

There are different categories of SSL algorithms. In this paper, we focus on {\bf contrastive learning}, a method that learns representations by distinguishing between similar and dissimilar data points. Specifically, the goal of contrastive learning is to map data into a feature space where similar sample pairs (positives) are pulled closer, while dissimilar ones (negatives) are pushed apart. This approach enables the model to capture essential patterns without requiring explicit feature engineering, shown in Figure~\ref{fig:contrastive_learning}.

Compared to traditional supervised learning, where models rely on labeled data, SSL learns directly from the intrinsic structure of the dataset without the reliance on expert knowledge.


\subsection{Neural Architecture Search (NAS)}
\label{sec:NASREL}

Neural Architecture Search (NAS) refers to techniques that automate the design of deep neural networks (DNNs) by searching for optimal architectures within a predefined space. Unlike manually designed AI models, NAS explores a large set of possible network architectures to identify configurations that yield superior performance, while also balancing the computational efficiency and model complexity. Typically, the NAS framework consists of three key components:

\begin{itemize}
    \item {\bf Search Space}: Defines the set of possible architectures, including choices for layer types, connections, and other hyperparameters.
    \item {\bf Search Strategy}: Defines how architectures are explored within the search space.
    \item {\bf Performance Estimation}: Evaluates the quality of candidate architectures, provides guidelines for subsequent updating.
\end{itemize}

There are many different strategies to implement the NAS framework, existing works include but are not limited to reinforcement learning (RL)-based NAS~\cite{zhang2021adaptive},  evolutionary algorithm (EA)-based NAS~\cite{pan2021neural}, and gradient-based NAS~\cite{shu2022unifying}. While NAS has been widely applied in fields such as image classification and natural language processing, its adoption in hardware security and HT detection remains limited. We will demonstrate how NAS was integrated into our framework to enable adaptive HT detection, details are discussed in Section~\ref{sec:nas}.

\section{Proposed Methodology}
\label{sec:prop}




\subsection{Overview}
\label{sec:overview}

To address these challenges described in Section~\ref{sec:bgd}, we propose SAND, a self-supervised learning (SSL)-based framework enhanced with neural architecture search (NAS) for adaptive and efficient HT detection. Figure~\ref{fig:overview} presents an overview of the proposed workflow. The framework consists of an {\bf upstream encoder} and a {\bf downstream classifier}, with the implementation divided into the following primary tasks:

\begin{figure}[htp]
    \centering
    \vspace{-0.1in}
    \includegraphics[width= 1.03\linewidth]{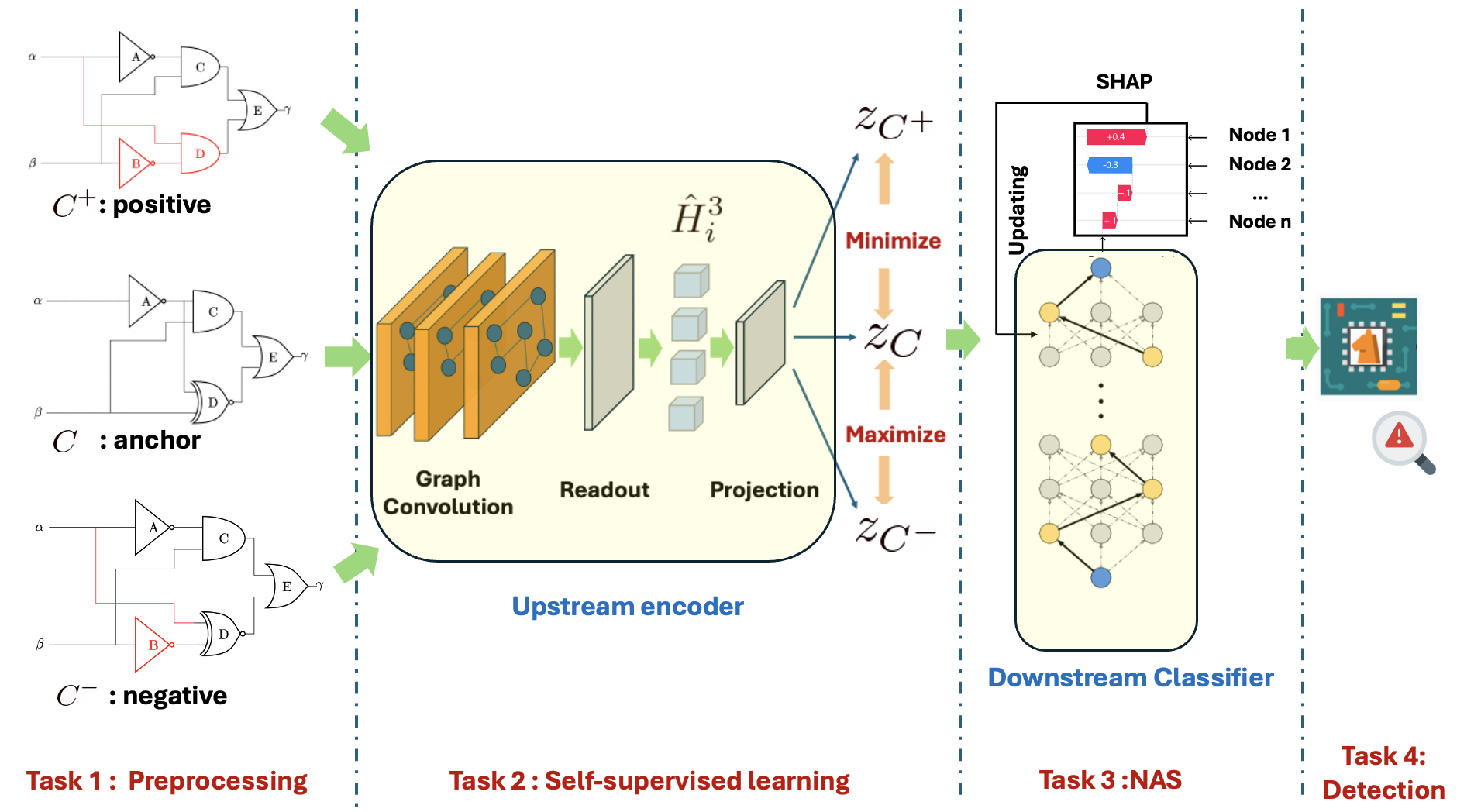}
    \vspace{-0.1in}
    \caption{The overview framework of SAND. The main module consists of an {\bf upstream encoder}, implemented through self-supervised learning (SSL) and a{\bf downstream classifier} crafted by neural architecture search (NAS).}
    \vspace{-0.2in}
    \label{fig:overview}
\end{figure}

\begin{itemize}
\item \textbf{Data preprocessing:} In this task, we implement the specific data preprocessing technique for hardware circuits, and further transform circuit information into a graph format for subsequent contrastive learning. (Section~\ref{sec:dataaug})

\item \textbf{Contrastive learning based SSL:} We adopt the contrastive learning approach to build a Graph Convolutional Network (GCN) as the upstream encoder, which learns the transferable feature representations of hardware circuits without relying on manually designed features. (Section~\ref{sec:gcnssl})  

\item \textbf{SHAP-based NAS:} The features are fed into the downstream classifier to enable adaptive HT detection. Specifically, we integrate NAS to automatically optimize its architecture, where Shapley value analysis (SHAP)\cite{winter2002shapley} is incorporated to refine the optimization process. (Section~\ref{sec:nas})  

\item \textbf{HT detection:} The finalized models are deployed in real-world HT detection tasks with enhanced adaptivity. 

\end{itemize}

\subsection{Data Preprocessing}
\label{sec:dataaug}



Effective contrastive learning relies on crafting both positive and negative samples. For hardware benchmarks, negative samples can be crafted by injecting Trojans into the anchor samples. \textit{However, generating positive samples for circuits is more challenging}. Directly using other benign circuits is impractical, as variations in scale and type lead to structural and functional disparities. Additionally, unlike image-based tasks, where simple augmentations like cropping or rotation preserve semantics, even a single gate modification in a circuit can alter its behavior. To address this, given a circuit $C$ as anchor input, we obtain its positive counterpart $C^+$ using the following transformations:

\begin{itemize}
    \item {\bf Logic-equivalent modification:} We introduce structural variations while maintaining logical equivalence. For example, using DeMorgan's theorem to replace an AND gate with a NAND gate followed by a NOT gate.  
    \item {\bf Subcircuit extraction:} We adopt the algorithm in ~\cite{vijaykrishnan1996subgen} to extract subcircuits from the given circuit, maintaining the partial structural patterns while reducing complexity.
    \item {\bf Relocation:} We also randomly alter the locations of independent sub-modules within the same benchmark, introducing structural variations while preserving functionality.
\end{itemize}


Meanwhile, we transform circuits into graph representations and apply Graph Convolutional Networks (GCNs) in the subsequent contrastive learning stage for effective feature embedding. Basically, we model the input circuit ${C}$ as a directed graph ${G(\mathcal{V,E})}$. Nodes $\mathcal{V}$ represent circuit cells, encoding their fundamental information (cell-type, fan-in/outs, etc.). Edges $\mathcal{E}$ denote connection wires between cells, capturing the circuit’s structural information.

\subsection{SSL-based Upstream Encoder}
\label{sec:gcnssl}
In this section, we describe how the generated data is leveraged within our SSL framework.
Specifically, we design the \textbf{upstream encoder} based on GCNs, followed by one readout layer, and one projection layer. The training is done by minimizing a \textit{hybrid contrastive loss function}. The entire framework consists of the following key components:

\subsubsection{Graph Convolution Layers (GCNs)}
\hfill\\
We adopt a \textit{three-layer} GCN to capture hierarchical information from circuit graphs. Given the graph ${G}$ as defined in the previous section, we assume each node $v \in \mathcal{V}$ has the initial attribute vector $\mathbf{h}_v^{(0)}$ encoding its fundamental information. Then the forward propagation in each GCN layer is defined as:

\begin{equation}
\mathbf{H}^{(l+1)} = \sigma\left( \tilde{\mathbf{D}}^{-\frac{1}{2}} \tilde{\mathbf{A}} \tilde{\mathbf{D}}^{-\frac{1}{2}} \mathbf{H}^{(l)} \mathbf{W}^{(l)} \right),
\end{equation}

where $\tilde{\mathbf{A}} = \mathbf{A} + \mathbf{I}$ is the adjacency matrix with self-loops, $\tilde{\mathbf{D}}$ is the corresponding degree matrix. The term $\tilde{\mathbf{D}}^{-\frac{1}{2}} \tilde{\mathbf{A}} \tilde{\mathbf{D}}^{-\frac{1}{2}}$ normalizes the adjacency matrix by the degree matrix. This ensures that the information passed between nodes is scaled by their degree, allowing the network to focus on the relative importance of neighbors.  
This graph convolution operation can be seen as a message-passing operation, where each node aggregates information from its neighbors (and itself).
At each iteration of graph convolution, each node embedding $\mathbf{h}_v^{(L)}$ captures both local connectivity and hierarchical circuit patterns, while propagating to the output layer to formalize a valid feature encoding for the given circuit. 

\subsubsection{Readout Layer}
\hfill\\
To obtain a standard representation among circuits with different scales, we apply a readout function aggregating node embeddings into a single vector using \textit{global sum pooling:}

\begin{equation}
\mathbf{\hat{H}^{(l)}} = \sum_{v \in \mathcal{V}} \mathbf{h}_v^{(L)}
\end{equation}


\subsubsection{Projection Layer}
\hfill\\
We also apply a projection layer to map the aggregated feature into a fixed-length vector using a fully connected dense layer:
\begin{equation}
\mathbf{z} = \sigma(\mathbf{W}_p \mathbf{\hat{H}}^{(l)} + b_p)
\end{equation}
Where \( \sigma \) is the non-linear activation function, $\mathbf{W}_p$ is the weight matrix of the projection layer, and $b_p$ is the bias term. The projection allows flexibility in adapting to different problem sizes.







\subsubsection{Hybrid Contrastive Loss Formulation}
\hfill\\
Once the projected feature vectors are obtained, we can train our proposed model to enable contrastive learning. To effectively learn meaningful representations for HT detection, we design a \textbf{hybrid contrastive loss} consisting of three complementary components:  
\begin{itemize}
    \item a \textit{Positive Contrastive Loss} to separate anchor circuits from the positive samples
    \item a \textit{Negative Contrastive Loss} to separate anchor circuits from the negative samples.
    \item a \textit{Global Clustering Loss} to encourage circuits of the same category to form well-separated clusters.  
\end{itemize}

\paragraph{\underline{Positive Contrastive Loss}}
The positive contrastive loss aims to minimize the distance between the anchor sample $C$ and all its corresponding positive samples $C^+_i$:
$\mathcal{L}_{P} = \sum_{i} \| z_C - z_{C^+_i} \|^2$.


\paragraph{\underline{Negative Contrastive Loss}}
Similarly, the negative contrastive loss is defined as:
$\mathcal{L}_{N} = \sum_{i} \max(0, m - \| z_C - z_{C^-_i}  \|^2)$,

note that $m$ is a margin hyperparameter that ensures a minimum distance between an original circuit and its Trojan-inserted variant. 


\paragraph{\underline{Global Clustering Loss}}

\begin{figure}[htp]
    \centering
     \vspace{-0.1in}
    \includegraphics[width= \linewidth]{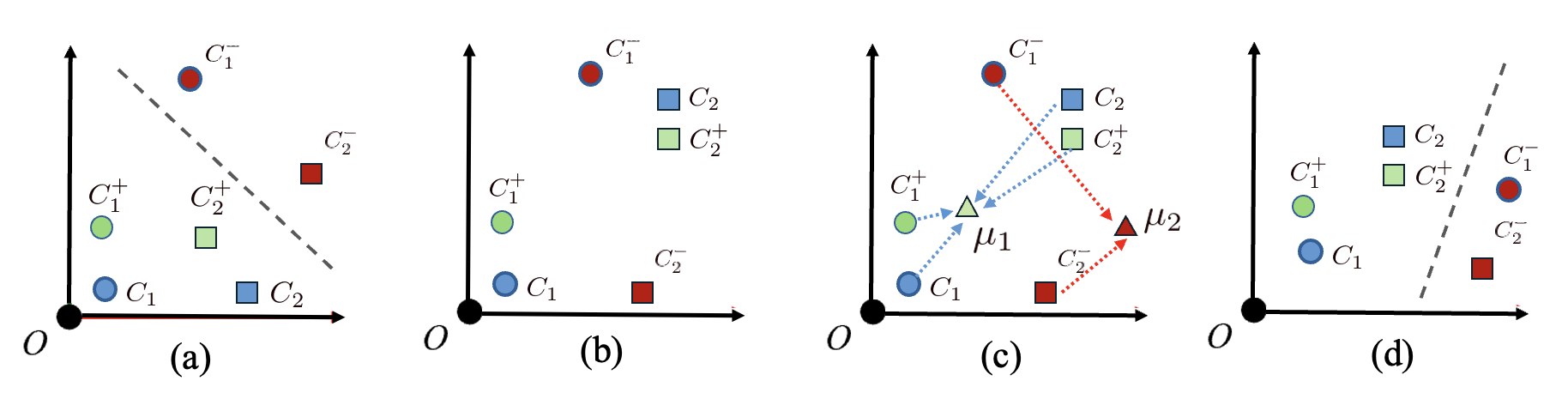}
    \vspace{-0.2in}
    \caption{Illustration of global clustering. Two benign circuits ($C_1$, $C_2$) and their associated positive/negative samples are shown. (a) Without global structure, contrastive loss ($\mathcal{L}_P$, $\mathcal{L}_N$) may succeed if $C_1$ and $C_2$ are close, (b) but fails when they are far apart. (c) Global clustering loss pulls samples toward their class centroid ($\mu_k$), (d) enhancing inter-class separation and reducing intra-class variance.}
    \vspace{-0.1in}
    \label{fig:centroid}
\end{figure}

For conventional contrastive learning, $\mathcal{L}_{P}$ and $ \mathcal{L}_{N}$ suffice to define the total loss. However, in HT detection, circuits within the same category (benign/malicious) can exhibit significant variations in structural composition and basic functionality. Moreover, the structural difference between an original circuit and its Trojan-injected variant is typically much smaller than that between it and another irrelevant benign circuit. As a result, if we merely rely on positive/negative contrastive loss, training an effective classifier becomes extremely challenging. Figure~\ref{fig:centroid}(b) illustrates a representative example of this phenomenon.  

To address this challenge, we introduce a \textit{global clustering loss} into our framework, which enforces category-level coherence by softly grouping circuits of the same class. 
Specifically, in addition to preserving pairwise relationships, we define class-specific centroids $\mu_k$ and encourage each circuit embedding to be close to its corresponding centroid, as shown in Figure~\ref{fig:centroid}(c). This ensures circuits of the same category (benign/malicious) form tight clusters in the feature space, balancing the intra-class variations:

\begin{equation}
\mathcal{L}_{\text{G}} = \sum_{C \in \mathcal{D}} \| z_C - \mu_k \|^2, \quad \text{where } \mu_k = \frac{1}{|\mathcal{D}_k|} \sum_{C \in \mathcal{D}_k} z_C
\end{equation}
where $\mathcal{D}_k$ denotes the set of all circuits belonging to class $k$, in our case $k\in\{0,1\}$ since HT detection is a binary classification. 

The final hybrid loss function integrates all three components, and we outline the complete training procedure in Algorithm~\ref{alg:ssl_training}. The encoder after training is expected to extract meaningful structural and functional representations of given circuits without a manual feature engineering step. 


\begin{algorithm}[htp]
    \caption{Contrastive Learning based SSL}
    \label{alg:ssl_training}
    \KwIn{Circuit graphs ${G} = \{G_i\}$, encoder parameters $\theta$}
    \KwOut{Trained upstream encoder}


    \BlankLine
    \tcc{Graph Convolution}
    \For{each epoch}{
        \For{each mini-batch}{
            \For{each GCN layer $l=1,2,3$}{
                   $ H^{(l+1)} = \sigma( \tilde{D}^{-\frac{1}{2}} \tilde{A} \tilde{D}^{-\frac{1}{2}} H^{(l)} W^{(l)} )$
            }
            
            \BlankLine
                $\hat{H}^{(L)} = \sum_{v \in \mathcal{V}} h_v^{(L)}$
                \BlankLine
                $z = \sigma(W_p \hat{H}^{(L)} + b_p)$
        }

    \BlankLine
    \tcc{Loss Computation and Backpropagation}

    $\quad \mathcal{L}_P = \sum_{i,j} \| z_{C_i} - z_{C_{ij}^+} \|^2$
    
    $\quad \mathcal{L}_{N} = \sum_{i,j} \max(0, m - \| z_{C_i} - - z_{C_{ij}^-}\|^2 )$
    
    $\quad \mathcal{L}_{\text{G}} = \sum_i \| z_{C_i} - \mu_k \|^2$    
    \BlankLine

    
    

    $\,\,$Total loss:
    $\quad \mathcal{L} = \lambda_1 \mathcal{L}_{P} + \lambda_2 \mathcal{L}_{N} + \lambda_3 \mathcal{L}_{\text{G}}$
    
    $\,\,$Backpropagation: $\theta = \theta - \nabla_\theta \mathcal{L}$ \;
    }
    \BlankLine
    \Return ${\theta}$ \;
\end{algorithm}



\subsection{NAS-based Downstream Classifier}
\label{sec:nas}


In this section, we describe how the upstream encoder’s outputs are utilized to train a downstream classifier for HT detection.

As discussed in Section~\ref{sec:MLHT}, another key challenge in ML-based HT detection is the limited adaptivity across circuit and Trojan variants. Manually designed architecture with fixed hyperparameters often lacks flexibility, leading to expensive retraining costs.

To address this issue, we employ Neural Architecture Search (NAS) to identify the optimal network architecture. By leveraging NAS, we aim to dynamically construct the classifier's architecture tailored to a focused HT detection task. Specifically, we apply the one-shot NAS training strategy consisting of two major steps: (i) SuperNet Training, and (ii) SHAP-based Pruning.

\subsubsection{SuperNet Training}
We begin with a large, over-parameterized network (\textit{SuperNet}) that encompasses all potential architectural configurations. At each layer, the SuperNet incorporates a diverse set of network components, including fully connected layers, convolutional layers, pooling layers, and activation functions. We first train this SuperNet to establish its core classification capability.

\subsubsection{SHAP-based Pruning} 
\begin{figure}[htp]
    \centering
     \vspace{-0.15in}
    \includegraphics[width= 0.9\linewidth]{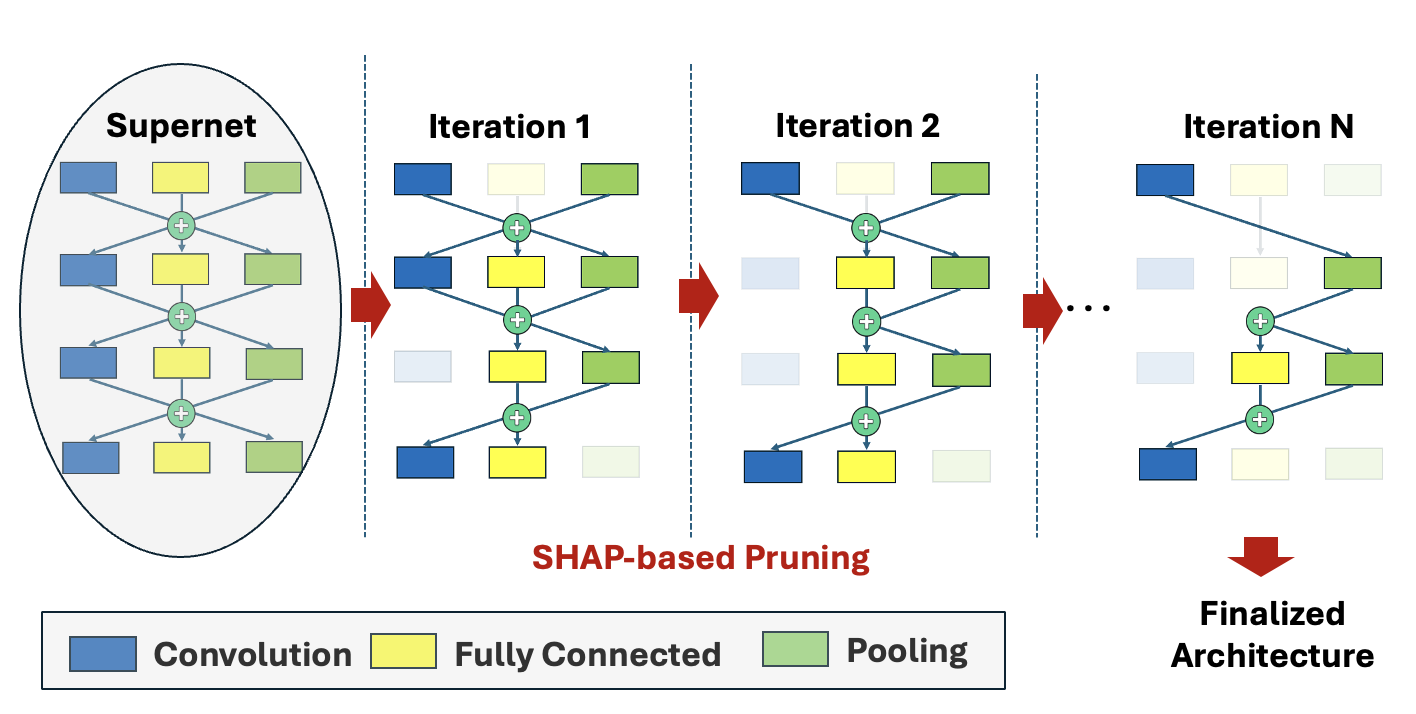}
    \vspace{-0.15in}
    \caption{The workflow of SHAP-based NAS for optimizing the downstream classifier architecture. The process begins with an over-parameterized SuperNet, components with low SHAP values are progressively pruned (faded). }
    \vspace{-0.15in}
    \label{fig:SHAPNAS}
\end{figure}

The next step is to apply evaluation metrics to identify the most effective components inside the SuperNet, extracting them as a streamlined sub-network optimized for the current HT detection task. We achieve this by identifying and pruning redundant components using Shapley value analysis (SHAP)~\cite{winter2002shapley}, which assigns each component an importance value based on its contribution to a model’s output. Higher SHAP values suggest that a component plays a crucial role in classification, while lower values indicate redundancy and will be pruned, leading to a more compact and task-specific model while preserving classification performance. Figure~\ref{fig:SHAPNAS} demonstrates the NAS process. 

\section{Experimental Evaluation}
\label{sec:exp}



\subsection{Experiment Setup}
\subsubsection{Implementation:}
To enable a comprehensive evaluation of proposed framework, the experimental is conducted on a host machine with an Intel i7 3.70GHz CPU, 32 GB RAM, and an NVIDIA RTX 3090 GPU. The proposed HT detection framework is implemented in Python, with model training and evaluation using the PyTorch deep learning library. Circuit netlists are transformed into graph representations using CircuitGraph~\cite{sweeney2020circuitgraph}, while SHAP-based NAS is implemented with libraries from~\cite{NIPS2017_7062}.

\subsubsection{Dateset:} 
We use a diverse of benchmarks to ensure comprehensive evaluation:
\begin{itemize}
    \item ISCAS85 \& ISCAS89~\cite{ISCAS89}: Standard combinational and sequential benchmark suites.
    \item Trust-Hub~\cite{THub}: Open repository with HT-infested designs across various circuit types.
    \item Open Designs~\cite{waterman2011risc}: Open designs based on RISC-V and MIPS architectures.
    \item Synthetic: Custom circuits with varied structures to simulate different attack scenarios.
\end{itemize}
Trojans are crafted and injected into these benchmarks using the tool provided from ~\cite{cruz2018automated}.

\subsubsection{Methods:}
In this paper, we consider and compare the performance of the following four ML-based HT detection schemes:
\begin{itemize}
    \item SVM~\cite{gubbi2023hardware}: Support vector machine (SVM) based method. 
    \item AdaTest~\cite{chen2023adatest}: Reinforcement learning based adaptive method.
    \item GATE-Net~\cite{wan2021contrastive}: State-of-the-art deep learning based method.
    \item SAND: Our proposed SSL-based and NAS-driven adaptive HT detection framework.
\end{itemize}





\subsection{Case Study: Detection Accuracy}

\begin{table*}[htp]
\footnotesize
    \centering
    \caption{Performance Comparison using accuracy (Acc), recall (Rec), precision (Pre) and F1 score (F1).}
    \label{tab:results}
    \vspace{-0.1in}
\begin{tabular}{|c|cccc|cccc|cccc|cccc|}
\hline
                 & \multicolumn{4}{c|}{SVM~\cite{gubbi2023hardware}}                                                         & \multicolumn{4}{c|}{AdaTest~\cite{chen2023adatest}}                                                      & \multicolumn{4}{c|}{GATE-Net~\cite{wan2021contrastive}}                                                  & \multicolumn{4}{c|}{SAND (Proposed Approach)}                                                                                  \\ \hline
\textbf{Bench}   & \multicolumn{1}{c|}{\textbf{Acc}}     & \multicolumn{1}{c|}{\textbf{Rec}}  & \multicolumn{1}{c|}{\textbf{Pre}}  & \textbf{F1}   & \multicolumn{1}{c|}{\textbf{Acc}}    & \multicolumn{1}{c|}{\textbf{Rec}}  & \multicolumn{1}{c|}{\textbf{Pre}}  & \textbf{F1}   & \multicolumn{1}{c|}{\textbf{Acc}}    & \multicolumn{1}{c|}{\textbf{Rec}}  & \multicolumn{1}{c|}{\textbf{Prec}} & \textbf{F1}   & \multicolumn{1}{c|}{\textbf{Acc}}    & \multicolumn{1}{c|}{\textbf{Rec}}  & \multicolumn{1}{c|}{\textbf{Pre}}  & \textbf{F1}   \\ \hline
c2670            & \multicolumn{1}{c|}{100.0\%}          & \multicolumn{1}{c|}{1.0}           & \multicolumn{1}{c|}{1.0}           & 1.0           & \multicolumn{1}{c|}{93.2\%}          & \multicolumn{1}{c|}{0.94}          & \multicolumn{1}{c|}{0.90}          & 0.92          & \multicolumn{1}{c|}{100.0\%}         & \multicolumn{1}{c|}{0.94}          & \multicolumn{1}{c|}{0.97}          & 0.96          & \multicolumn{1}{c|}{100.0\%}         & \multicolumn{1}{c|}{1.0}           & \multicolumn{1}{c|}{1.0}           & 1.0           \\ \hline
c5315            & \multicolumn{1}{c|}{100.0\%}          & \multicolumn{1}{c|}{1.0}           & \multicolumn{1}{c|}{1.0}           & 1.0           & \multicolumn{1}{c|}{97.6\%}          & \multicolumn{1}{c|}{0.99}          & \multicolumn{1}{c|}{0.95}          & 0.96          & \multicolumn{1}{c|}{100.0\%}         & \multicolumn{1}{c|}{0.91}          & \multicolumn{1}{c|}{0.92}          & 0.92          & \multicolumn{1}{c|}{100.0\%}         & \multicolumn{1}{c|}{1.0}           & \multicolumn{1}{c|}{1.0}           & 1.0           \\ \hline
c6288            & \multicolumn{1}{c|}{88.5\%}           & \multicolumn{1}{c|}{0.88}          & \multicolumn{1}{c|}{0.83}          & 0.84          & \multicolumn{1}{c|}{94.5\%}          & \multicolumn{1}{c|}{0.98}          & \multicolumn{1}{c|}{0.89}          & 0.93          & \multicolumn{1}{c|}{98.1\%}          & \multicolumn{1}{c|}{0.85}          & \multicolumn{1}{c|}{0.89}          & 0.87          & \multicolumn{1}{c|}{100.0\%}         & \multicolumn{1}{c|}{0.99}          & \multicolumn{1}{c|}{0.99}          & 0.99          \\ \hline
c7552            & \multicolumn{1}{c|}{87.2\%}           & \multicolumn{1}{c|}{0.81}          & \multicolumn{1}{c|}{0.92}          & 0.86          & \multicolumn{1}{c|}{84.9\%}          & \multicolumn{1}{c|}{0.86}          & \multicolumn{1}{c|}{0.81}          & 0.83          & \multicolumn{1}{c|}{100.0\%}         & \multicolumn{1}{c|}{0.89}          & \multicolumn{1}{c|}{0.91}          & 0.90          & \multicolumn{1}{c|}{100.0\%}         & \multicolumn{1}{c|}{1.0}           & \multicolumn{1}{c|}{1.0}           & 1.0           \\ \hline
s13207           & \multicolumn{1}{c|}{78.5\%}           & \multicolumn{1}{c|}{0.77}          & \multicolumn{1}{c|}{0.79}          & 0.78          & \multicolumn{1}{c|}{90.0\%}          & \multicolumn{1}{c|}{0.92}          & \multicolumn{1}{c|}{0.89}          & 0.90          & \multicolumn{1}{c|}{95.7\%}          & \multicolumn{1}{c|}{0.94}          & \multicolumn{1}{c|}{0.95}          & 0.95          & \multicolumn{1}{c|}{100.0\%}         & \multicolumn{1}{c|}{1.0}           & \multicolumn{1}{c|}{1.0}           & 1.0           \\ \hline
s15850           & \multicolumn{1}{c|}{78.8\%}           & \multicolumn{1}{c|}{0.75}          & \multicolumn{1}{c|}{0.79}          & 0.77          & \multicolumn{1}{c|}{82.0\%}          & \multicolumn{1}{c|}{0.84}          & \multicolumn{1}{c|}{0.79}          & 0.81          & \multicolumn{1}{c|}{90.0\%}          & \multicolumn{1}{c|}{0.84}          & \multicolumn{1}{c|}{0.97}          & 0.90          & \multicolumn{1}{c|}{98.8\%}          & \multicolumn{1}{c|}{0.99}          & \multicolumn{1}{c|}{0.99}          & 0.99          \\ \hline
s35932           & \multicolumn{1}{c|}{73.1\%}           & \multicolumn{1}{c|}{0.80}          & \multicolumn{1}{c|}{0.70}          & 0.74          & \multicolumn{1}{c|}{79.5\%}          & \multicolumn{1}{c|}{0.80}          & \multicolumn{1}{c|}{0.77}          & 0.78          & \multicolumn{1}{c|}{80.5\%}          & \multicolumn{1}{c|}{0.75}          & \multicolumn{1}{c|}{0.86}          & 0.80          & \multicolumn{1}{c|}{98.8\%}          & \multicolumn{1}{c|}{0.97}          & \multicolumn{1}{c|}{0.99}          & 0.98          \\ \hline
AES-T100         & \multicolumn{1}{c|}{85.9\%}           & \multicolumn{1}{c|}{0.84}          & \multicolumn{1}{c|}{0.86}          & 0.85          & \multicolumn{1}{c|}{100.0\%}         & \multicolumn{1}{c|}{1.0}           & \multicolumn{1}{c|}{1.0}           & 1.0           & \multicolumn{1}{c|}{96.5\%}          & \multicolumn{1}{c|}{0.96}          & \multicolumn{1}{c|}{0.96}          & 0.96          & \multicolumn{1}{c|}{100.0\%}         & \multicolumn{1}{c|}{1.0}           & \multicolumn{1}{c|}{1.0}           & 1.0           \\ \hline
AES-T200         & \multicolumn{1}{c|}{79.3\%}           & \multicolumn{1}{c|}{0.77}          & \multicolumn{1}{c|}{0.80}          & 0.78          & \multicolumn{1}{c|}{96.2\%}          & \multicolumn{1}{c|}{0.97}          & \multicolumn{1}{c|}{0.95}          & 0.96          & \multicolumn{1}{c|}{95.1\%}          & \multicolumn{1}{c|}{0.92}          & \multicolumn{1}{c|}{0.97}          & 0.94          & \multicolumn{1}{c|}{96.4\%}          & \multicolumn{1}{c|}{1.0}           & \multicolumn{1}{c|}{1.0}           & 1.0           \\ \hline
AES-T1000        & \multicolumn{1}{c|}{86.2\%}           & \multicolumn{1}{c|}{0.86}          & \multicolumn{1}{c|}{0.87}          & 0.86          & \multicolumn{1}{c|}{90.5\%}          & \multicolumn{1}{c|}{0.93}          & \multicolumn{1}{c|}{0.88}          & 0.90          & \multicolumn{1}{c|}{90.4\%}          & \multicolumn{1}{c|}{0.85}          & \multicolumn{1}{c|}{0.95}          & 0.90          & \multicolumn{1}{c|}{98.8\%}          & \multicolumn{1}{c|}{1.0}           & \multicolumn{1}{c|}{1.0}           & 1.0           \\ \hline
MIPS             & \multicolumn{1}{c|}{66.7\%}           & \multicolumn{1}{c|}{0.69}          & \multicolumn{1}{c|}{0.64}          & 0.66          & \multicolumn{1}{c|}{88.8\%}          & \multicolumn{1}{c|}{0.89}          & \multicolumn{1}{c|}{0.87}          & 0.88          & \multicolumn{1}{c|}{73.5\%}          & \multicolumn{1}{c|}{0.72}          & \multicolumn{1}{c|}{0.75}          & 0.74          & \multicolumn{1}{c|}{94.0\%}          & \multicolumn{1}{c|}{0.95}          & \multicolumn{1}{c|}{0.93}          & 0.94          \\ \hline
RISCV            & \multicolumn{1}{c|}{50.8\%}           & \multicolumn{1}{c|}{0.53}          & \multicolumn{1}{c|}{0.50}          & 0.51          & \multicolumn{1}{c|}{82.0\%}          & \multicolumn{1}{c|}{0.85}          & \multicolumn{1}{c|}{0.80}          & 0.83          & \multicolumn{1}{c|}{77.2\%}          & \multicolumn{1}{c|}{0.77}          & \multicolumn{1}{c|}{0.77}          & 0.77          & \multicolumn{1}{c|}{90.4\%}          & \multicolumn{1}{c|}{0.89}          & \multicolumn{1}{c|}{0.92}          & 0.91          \\ \hline
Synthetic        & \multicolumn{1}{c|}{73.2\%}           & \multicolumn{1}{c|}{0.70}          & \multicolumn{1}{c|}{0.75}          & 0.72          & \multicolumn{1}{c|}{85.4\%}          & \multicolumn{1}{c|}{0.88}          & \multicolumn{1}{c|}{0.80}          & 0.84          & \multicolumn{1}{c|}{76.9\%}          & \multicolumn{1}{c|}{0.74}          & \multicolumn{1}{c|}{0.79}          & 0.77          & \multicolumn{1}{c|}{91.0\%}          & \multicolumn{1}{c|}{0.90}          & \multicolumn{1}{c|}{0.91}          & 0.91          \\ \hline
\textbf{Average} & \multicolumn{1}{c|}{\textbf{81.3 \%}} & \multicolumn{1}{c|}{\textbf{0.80}} & \multicolumn{1}{c|}{\textbf{0.80}} & \textbf{0.80} & \multicolumn{1}{c|}{\textbf{90.6\%}} & \multicolumn{1}{c|}{\textbf{0.91}} & \multicolumn{1}{c|}{\textbf{0.88}} & \textbf{0.90} & \multicolumn{1}{c|}{\textbf{92.2\%}} & \multicolumn{1}{c|}{\textbf{0.89}} & \multicolumn{1}{c|}{\textbf{0.92}} & \textbf{0.90} & \multicolumn{1}{c|}{\textbf{98.1\%}} & \multicolumn{1}{c|}{\textbf{0.98}} & \multicolumn{1}{c|}{\textbf{0.98}} & \textbf{0.98} \\ \hline
\end{tabular}
\end{table*}

Table~\ref{tab:results} compares the performance of our proposed method, SAND, with other approaches for HT detection using accuracy (Acc), recall (Rec), precision (Pre), and F1-score (F1) across various benchmarks. Each row corresponds to a benchmark, with average values summarized in the last row.

SVM exhibits the lowest overall performance, as expected due to its lightweight architecture and limited capacity, which leads to overfitting and poor generalization for large or synthetic benchmarks. For example, despite {\bf 100\%} accuracy on c2670 and c5315, it drops to {\bf 50.8\%} on RISC-V designs, comparable to random guessing. AdaTest, a reinforcement learning-based method, improves over SVM with more consistent results. However, its low precision and high recall suggest a tendency to over-detect Trojans, majorly due to its adaptive feedback-driven strategy. This behavior helps catch unseen threats but also raises false positives by misclassifying benign instances. As a comparison, GATE-Net, based on deep learning, achieves stronger performance than these two methods, but slightly degrades on larger benchmarks, indicating scalability limitations. It also shows higher precision and lower recall, reflecting a more conservative prediction style that aims to avoid false alarms but ends up missing threats.
In contrast, SAND utilizes SSL to outperform the others with up to {\bf 18.3\%} gain in detection accuracy and consistently strong performance across benchmarks.

\subsection{Case Study: SSL-based Feature Embedding}

\begin{figure}[htp]
    \centering
     \vspace{-0.15 in}
    \includegraphics[width= \linewidth]{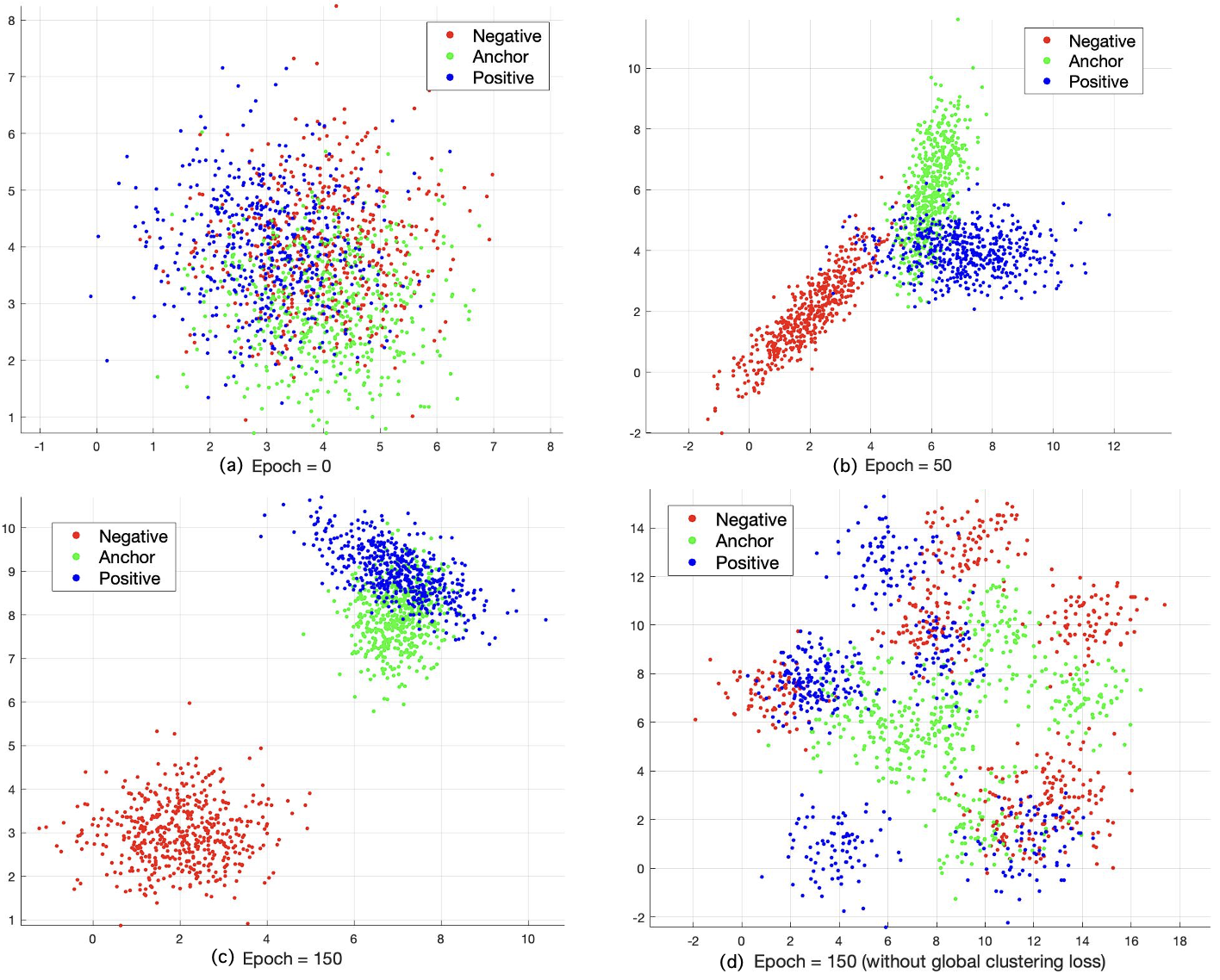}
    \vspace{-0.15 in}
    \caption{2D feature embeddings learned by SAND. (a) Epoch 0: Features are randomly scattered. (b) Epoch 50: Early clustering with some category overlap. (c) Epoch 150: Clear, well-separated clusters. (d) Without global clustering loss: fragmented intra-class clusters lead to intermixing.}
    \vspace{-0.2 in}
    \label{fig:scatterplot}
\end{figure}

To evaluate the effectiveness of the proposed SSL framework for automatic feature encoding, we analyze the evolution of feature embeddings over training epochs visualized by 2-D Principal Component Analysis (PCA). Figure~\ref{fig:scatterplot} illustrates the incremental learning process. An ablation test of the global clustering loss is also performed to address its necessity for HT detection.

At the initial stage (Epoch = 0, Figure~\ref{fig:scatterplot}(a)), the initialized values are scattered randomly in space. As training progresses to Epoch = 50 (Figure~\ref{fig:scatterplot}(b)), initial clusters begin to emerge, but the distances between clusters are not sufficiently far. By Epoch = 150 (Figure~\ref{fig:scatterplot}(c)), the feature embeddings are well-formed, malicious samples are clustered together and pushed far from the benign and positive samples, making it a valid feature encoding. This result highlights the effectiveness of the proposed SSL framework, where meaningful feature representations emerge automatically through the learning process, without requiring predefined expert-driven heuristics or manual feature selection. 

Notably, if we disable the global clustering loss and solely rely on contrastive loss, the learned representations exhibit a drastic difference (Figure~\ref{fig:scatterplot}(d)). As we can see, instead of forming compact clusters per category, multiple local clusters exist within each category, leading to intermixing between different categories, reducing the model's ability to generalize across diverse benchmarks.

\subsection{Case Study: NAS-based Adaptability}
\label{sec:NASadapt}

\begin{figure}[htp]
    \centering
     \vspace{-0.1 in}
    \includegraphics[width= \linewidth]{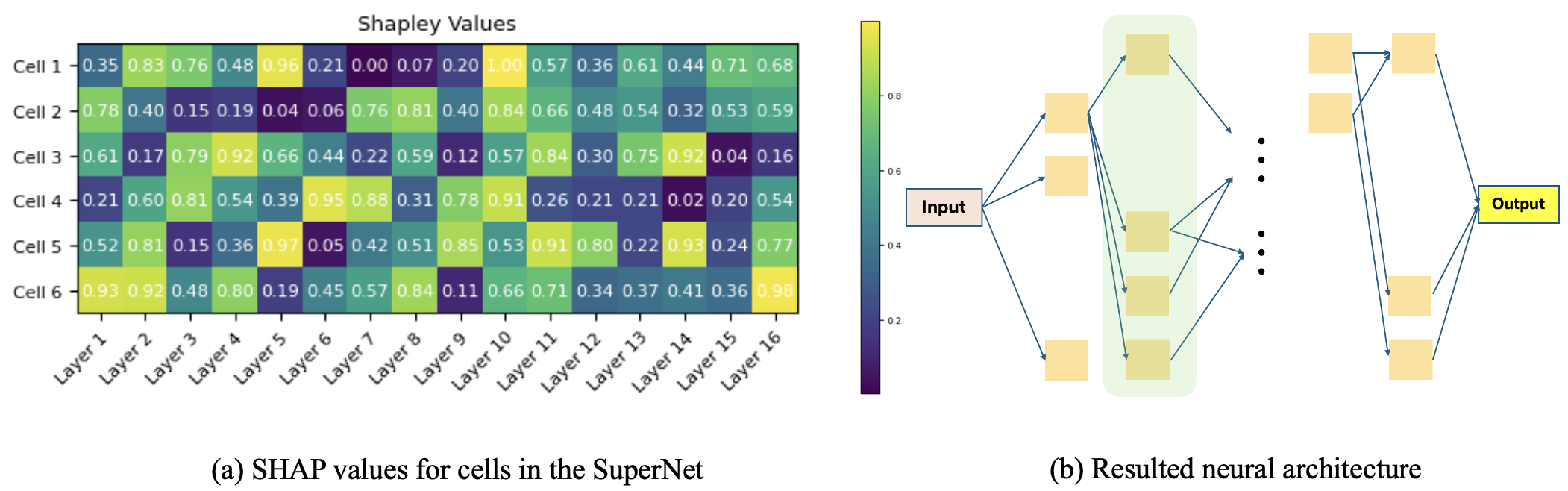}
    \vspace{-0.2 in}
    \caption{The automatically generated model architecture through SHAP-based NAS.}
    \vspace{-0.15 in}
    \label{fig:NASEXP}
\end{figure}

In this section, we evaluate the performance of our NAS-based framework. First, our framework enables automated model design by analyzing the contributions of different cell types (e.g., convolution, fully connected, pooling) within a 16-layer SuperNet, where each layer contains 6 cells. As shown in Figure~\ref{fig:NASEXP}(a). 

By computing the 2D SHAP values for all cells and applying thresholding, the final architecture in our experiment is shown in Figure~\ref{fig:NASEXP}(b). While seemingly intuitive, this architecture was obtained automatically by NAS, without the need for expert-driven feature engineering or heuristic assumptions.

\begin{figure}[htp]
    \centering
    \vspace{-0.15 in}
    \includegraphics[width=0.7\linewidth]{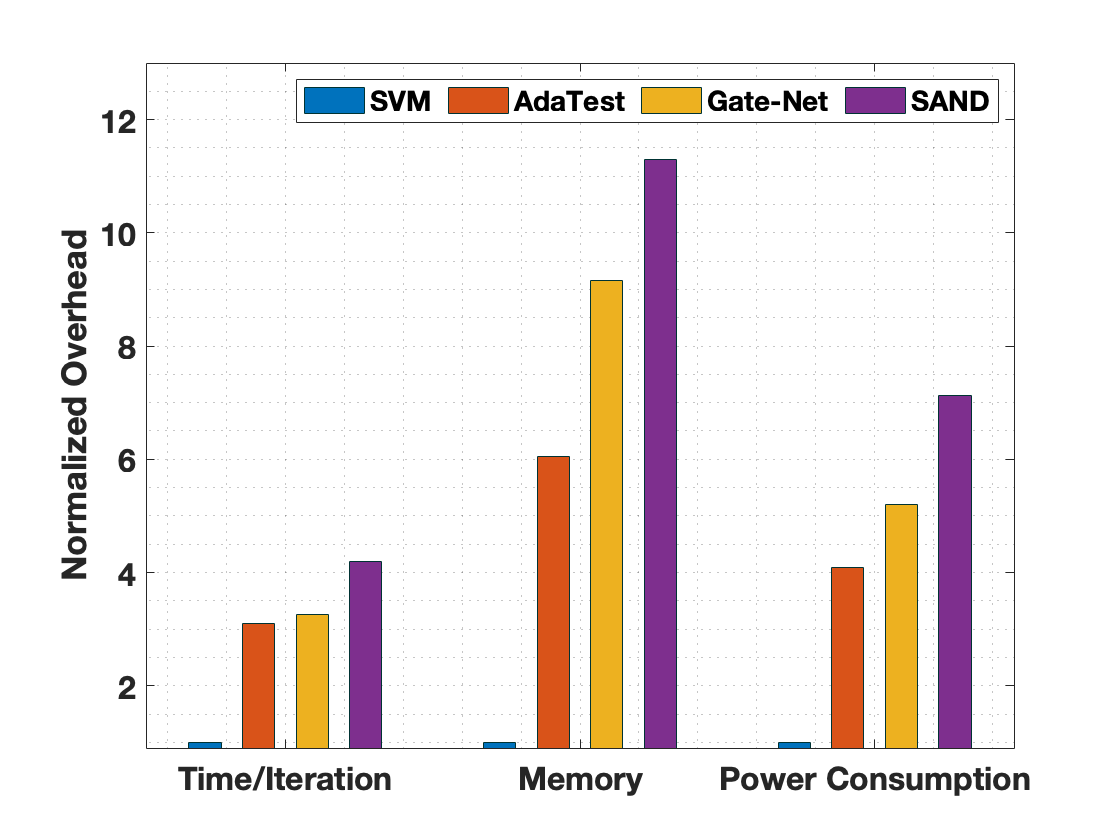}
    \vspace{-0.15 in}
    \caption{Training overhead comparison across methods.}
    \vspace{-0.15 in}
    \label{fig:hist2}
\end{figure}

However, the NAS framework also incur higher overhead compared to the baseline approaches, shown in Figure~\ref{fig:hist2}. Specifically, SAND introduces increased training time and memory usage due to the need to explore and maintain a supernet during architecture optimization, and also exhibits elevated power consumption during training. Despite this additional cost, the NAS-based architecture  
grants strong adaptability to our proposed method. For example, we have conducted experiments using unseen benchmarks to assess performance degradation and retraining overhead. The results are shown in Table~\ref{tab:adaptability}. 

As we can see, while traditional methods experience significant drops in accuracy, our approach retain strong performance. This robustness can be attributed to the utilization of self-supervised learning (SSL), which captures fundamental netlist properties without relying on manually crafted features, further improving adaptability. 
Additionally, the proposed approach minimizes retraining overhead, since we only need to fine-tune the downstream classifier encoded within the SuperNet, making adaptation to new benchmarks efficient. In contrast, all the other methdos requires longer epochs, leading to higher total costs for retraining.

\begin{table}[htp]
\footnotesize
    \centering
    \vspace{-0.1 in}
    \caption{Performance of adaptive HT detection.\label{tab:adaptability}}
        \vspace{-0.1 in}
    \begin{tabular}{l|c|c|c|c}
        \hline
        \textbf{Method} & \textbf{Seen} & \textbf{Unseen} & \textbf{Perf-Drop} & \textbf{Retraining (Epochs)} \\
        \hline
        SVM~\cite{gubbi2023hardware} & 81.3\% & 65.2\% & -16.1\% & 38  \\
        AdaTest~\cite{chen2023adatest} & 90.6\% & 70.8\% & -18.3\% & 32 \\
        GATE-Net~\cite{wan2021contrastive} & 92.2\% & 75.4\% & -16.8\% & 22 \\
        SAND (Proposed) & 98.1\% & 94.2\% & -3.9\% & 7 \\
        \hline
    \end{tabular}
        \vspace{-0.2 in}
\end{table}

\vspace{-0.1 in}
\subsection{Case Study: Stability}
Stability is another crucial factor for real-world HT detection tasks. To evaluate the stability of our proposed method, we conducted 20 trials to compare the detection accuracy of all methods.

\begin{figure}[htp]
    \centering
     \vspace{-0.2 in}
    \includegraphics[width= 0.95\linewidth]{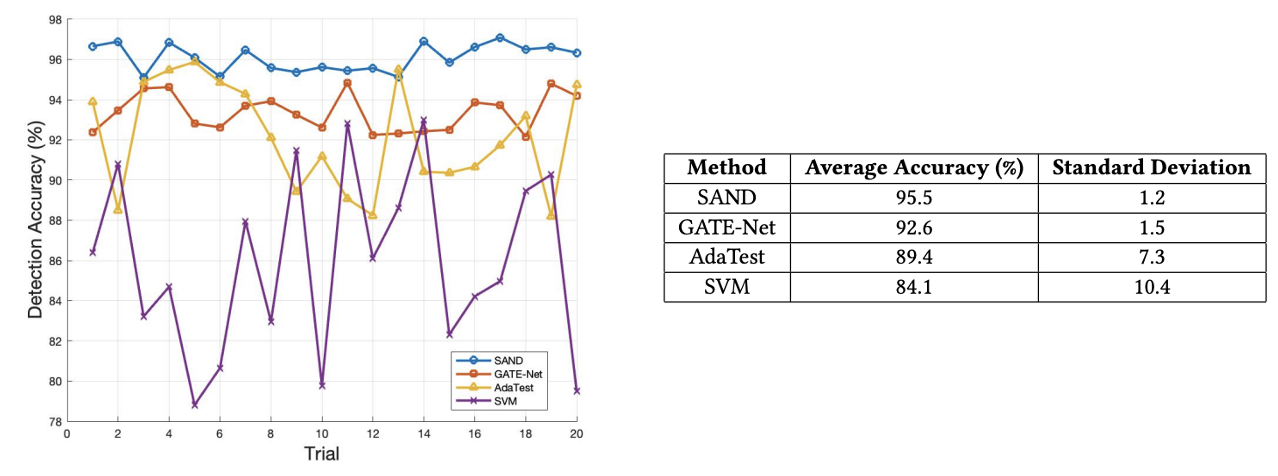}
    \vspace{-0.2 in}
    \caption{Detection accuracy comparison over 20 trials\label{fig:stability}}
    \vspace{-0.15 in}
\end{figure}

As shown in Figure~\ref{fig:stability}, we observe that both SAND and GATE-Net exhibit the highest stability, maintaining consistent and stable performance among multiple trials. In contrast, SVM demonstrates significant instability, frequently overfitting to specific patterns, resulting in large fluctuations in accuracy. AdaTest and GATE-NET shows a more stable performance compared to SVM, but still lags behind SAND. Our framework achieves superior stability due to its combination of three core design choices: (1) the data processing step ensures balanced supervision and mitigates bias from imbalanced datasets; (2) the global clustering loss enforces consistent inter-class separation, reducing variance in decision boundaries; and (3) the NAS-selected architecture is tailored to maintain performance across varying circuit complexities, minimizing performance fluctuation during deployment. 

\vspace{-0.05 in}
\section{Conclusion}
\label{sec:conclude}
Hardware Trojans (HTs) pose a critical threat to embedded systems. Existing ML-based HT detection methods suffer from reliance on ad-hoc feature engineering and limited adaptivity. To address this, this paper presents a self-supervised learning (SSL) and Neural Architecture Search (NAS) based framework for efficient HT detection. The proposed method integrates an upstream encoder trained with contrastive learning to automatically perform feature extraction, followed by an adaptive downstream classifier optimized by NAS for efficient deployment. Experimental results demonstrate that our approach significantly outperforms existing methods in detection accuracy, while maintaining decent adaptivity to unseen HT variants.

\bibliographystyle{ACM-Reference-Format}
\bibliography{sample-base}

\end{document}